# Spatiotemporal observation of quantum crystallization of electrons

Hideaki Murase, Shunto Arai, Tatsuo Hasegawa, Kazuya Miyagawa and Kazushi Kanoda*

*Department of Applied Physics, University of Tokyo, Bunkyo-ku, Tokyo 113-8656, Japan.*

*Corresponding author. Email: kanoda@ap.t.u-tokyo.ac.jp

**Liquids crystallize as they cool; however, when crystallization is avoided in some way, they supercool, maintaining their liquidity, and freezing into glass at low temperatures, as ubiquitously observed[1–3]. These metastable states crystallize over time through the classical dynamics of nucleation and growth[2,4,5]. However, it was recently found that Coulomb interacting electrons on charge-frustrated triangular lattices exhibit supercooled liquid and glass[6] with quantum nature[7] and they crystallize[8], raising fundamental issues : what features are universal to crystallization at large and specific to that of quantum systems? Here, we report our experimental challenges that address this issue through the spatiotemporal observation of electronic crystallization in an organic material. With Raman microspectroscopy, we are the first to successfully perform real-space and real-time imaging of electronic crystallization. The results directly capture strongly temperature-dependent crystallization profiles indicating that nucleation and growth proceed at distinctive temperature-dependent rates, which is common to conventional crystallization. Remarkably, however, the growth rate is many orders of magnitude larger than that in the conventional case, which is attributable to the quantum effect. The temperature characteristics of nucleation and growth are**

**universal, whereas unusually fast growth kinetics features quantum crystallization where a quantum-to-classical catastrophe occurs in interacting electrons.**

Crystallization from supercooled liquids or glasses, one of the most fundamental nonequilibrium phenomena, is widely observed in a range of condensed matter[2,4,5]. Conventionally, the phenomenon is governed by classical dynamics of the constituent elements and is known to proceed via two sequential processes, the nucleation of crystal seeds and their growth. If the constituent elements have a quantum nature, how does the quantum nature affect the crystallization process that has ever been captured by classical dynamics and thermodynamics? The recent discovery of electron glass and its crystallization in molecular materials has offered an experimental avenue to tackle this fundamental question in soft matter physics. Electron glass, which is called charge glass (CG), is exhibited by Coulomb-interacting electrons that fail to crystallize into a charge order (CO) on charge-frustrated lattices[9,10]. For example, on a triangular lattice that is half-filled (a band is quarter-filled) by electrons, they have difficulty in finding a stable CO owing to the large number of degenerate charge configurations; thus, they instead form CG[11–13]. Actually, the competitive appearance of CO and CG is demonstrated in the layered molecular conductors, θ-$(BEDT\text{-}TTF)_2X$ [X=$RbZn(SCN)_4$, $CsZn(SCN)_4$ and $TlCo(SCN)_4$], with quasi-triangular lattices of BEDT-TTF molecules (Figs. 1a, 1b), where either CO or CG appears depending on the anisotropy of the triangular lattice controlling the degree of charge frustration[6,14,15]. The CG shows the hallmarks of conventional classical glasses, such as slow fluctuations, medium-scale correlation and aging indicative of nonequilibrium[6,14,15]. Moreover, remarkably, the electrons in CG are found to have itinerant character by nuclear magnetic resonance (NMR) and transport measurements, suggesting that CG has a quantum-classical energetic hierarchy[7]. This may have relevance to a theoretically proposed quantum glass[16].

Among these compounds, θ-$(BEDT\text{-}TTF)_2RbZn(SCN)_4$ (θ-RbZn) is a key material staying on the verge between CO and CG[17]. θ-RbZn undergoes a phase transition to CO at $T_{CO}$=200 K when cooled slowly, e.g., at a rate of 1 K/min[18–20]. However, when it is rapidly cooled at 5 K/min or faster, CO gives way to a supercooled charge liquid (SCL) and SCL freezes to CG at low temperatures (Fig. 1c)[6,14,15]. The crystallization from SCL or CG to CO was previously investigated by transport and NMR measurements, which characterized the evolution of the CO volume fraction and suggested different crystallization profiles at high and low temperatures; these results were reminiscent of classical glasses[8,21]. Spatiotemporal observation of crystallization is expected to directly reveal the nucleation and growth process to possibly elucidate the quantum effect; however, this has never been done. Here, we exploit Raman microspectroscopy to visualize the electron crystallization process in real space and real time for the first time. The microscopic crystallization profile and the unusually rapid growth of crystal seeds not compatible with classical glass are presented.

Raman spectroscopy sensitively probes the molecular charge in organic conductors[22] by activating the C=C stretching modes in BEDT-TTF, which are known to be particularly charge-sensitive[23-26].In the present study, we used the samples with the central double-bonded carbons enriched by $^{13}$C isotopes (Fig. 2a), which make the stretching mode well separated from other modes[19]. The Raman spectrum of this mode distinguishes between CO and SCL/CG. Figure 2a shows the time evolution of the Raman spectra of the charge-sensitive C=C stretching mode ($\nu_2$ mode) measured in an area of ~$0.1\times0.1$ mm$^2$ after rapid cooling to 190 K. Just after that, an SCL is metastabilized so that the spectrum exhibits a broad peak indicative of a structureless nonuniform charge distribution. Then, it evolves into the two-peak structure of CO on the time scale of ~$10^3$ s, as seen in Fig. 2a. In the CO phase, the Raman spectrum has two peaks coming from charge-rich and charge-poor molecules, reproducing the previous results[11]; the different peak intensities originate from the different Raman tensors of charge-rich and charge-poor sites[27], possibly related to electron-molecular vibration coupling. The time evolution of the electronic crystallization is characterized by fitting the observed spectrum, $I(\nu, t)$, at a time, $t$, by a linear combination of the spectra of CO, $I^{CO}(\nu)$, and SCL/CG, $I^{CG}(\nu)$:

$$I(\nu, t) = A^{CO}(t)I^{CO}(\nu) + A^{CG}(t)I^{CG}(\nu), \qquad (1)$$

where $\nu$ is the Raman shift, and $A^{CO}$ and $A^{CG}$ are the fitting parameters reflecting the spectral weights (refer to the Supplementary Note I). Throughout crystallization, the experimental $I(\nu, t)$ is well fitted by $I^{CO}(\nu)$ and $I^{CG}(\nu)$, and the volume fraction of CO is:

$$\phi(t) = \frac{A^{CO}(t)}{A^{CO}(t) + A^{CG}(t)}. \qquad (2)$$

To globally characterize the temperature profile of CO evolution from SCL or CG, we examined the time evolution of $\phi(t)$ from the integrated Raman spectra over the areal scale of ~$0.1 \times 0.1$ mm$^2$ during isothermal crystallization at various temperatures. The sample that was initially kept at 210 K, which is above $T_{CO}$, was rapidly cooled at 30 K/min to $T_q$, at which the Raman spectra were recorded as a function of time. After that, the sample was warmed to 210 K, and the sequence was repeated for different $T_q$ values. Figures 2b and 2c show the time evolution of $\phi(t)$ at $T_q$ values above and below 165 K, respectively. When $T_q$ is above 165 K, $\phi(t)$ vanishes for a while ($4 \times 10^2$-$10^4$ s) and then increases rapidly once it starts to rise (Fig. 2b). This means that the nucleation of the crystal seeds requires a certain incubation time, which is shorter at lower $T_q$, and immediately after nucleation occurs, the seeds spatially extend. In contrast, for $T_q$ below 165 K, the rise of $\phi$ from zero conversely requires a longer time at lower $T_q$ and is not sharp, followed by a gradual increase over time (Fig. 2c). In this temperature range, the time evolution of $\phi(t)$ is well described by the Avrami equation[28]:

$$\phi(t) = 1 - \exp(-Kt^n), \qquad (3)$$

where $K$ is the rate constant and $n$ is the Avrami exponent. This means that crystallization proceeds via the nucleation of numerous microcrystals and their growth. The deduced Avrami exponent $n$ is in the range of 2.6-3.5, which suggests that crystal growth proceeds in two-dimensional planes since $n$ is related to the spatial dimension of growth $d$ by $n=d+1$ in ordinary cases. Figure 2d displays the contour plot of $\phi$ in the $t$-$T_q$ plane to construct the so-called time-temperature-transformation (TTT) diagram. The contour lines form nose structures characteristic of the nucleation and growth mechanism[2]. Remarkably, a nose temperature of ~165 K divides the high- and low-temperature regimes described above, consistent with previous transport and NMR studies[8,21].

To visualize the spaciotemporal evolution of the crystallization process, we utilized a Raman microspectroscopy technique with a spatial resolution of 6.5 μm, in essence, one pixel is 6.5$\times$6.5 μm$^2$. Since the time evolution of $\phi$ behaves differently above and below $T_n$ (=165 K), we performed Raman microspectroscopy experiments at $T_q$ =195 K (>$T_n$) and 155 K (<$T_n$) to comparatively examine the two regimes.

Figure 3a shows snapshots of the crystallization in real space at $T_q$ = 195 K, represented by the contour plot of $\phi$. For a certain time after quenching, the system is an SCL everywhere in the sample (the blue colour over the entire sample surface of 65$\times$130 μm$^2$); however, once a part of the SCL crystallizes to form a CO microdomain, it rapidly extends over the entire area. We confirm that the image does not practically evolve with time while taking one image in the present spatial and time resolutions (Supplementary Note VI). In contrast, the contour plot of $\phi$ at $T_q$ = 155 K shown in Fig. 3b indicates that the time evolution of $\phi$ occurs nearly homogeneously in the whole system. This means that the spatial scale of nucleation and growth is considerably smaller than the resolution of 6.5 μm. A statistical error analysis shows that the mottled colours in Fig. 3b are not due to the spatial fluctuation of CO microcrystal density but to the spectral noise (Supplementary Note VII).

It is known that the times required for nucleation and the growth speed have different temperature dependencies. In the supercooled liquid state at high temperatures, nucleation takes a long time; however, the nuclei that happen to arise grow fast. In the glass state below $T_n$, nucleation events occur everywhere in the sample before individual nuclei spatially grow because of the slow growth speed; thus, crystallization proceeds in the form of an increasing number of fine nucleation sites, which are averaged and blurred at a spatial resolution of 6.5 μm. In such a case, the time evolution of the averaged volume fraction of the crystal seeds is known to be described by the Avrami equation[28], which is actually followed by the experimental $\phi$ values (Fig. 2c).

Furthermore, we investigated the crystal growth speed by analysing the time evolution of the Raman images. The $\phi$ values along the arrow at different times and a temperature of 195 K in Fig. 3a are plotted in Fig. 4a and fitted by the phenomenological form[29]:

$$\phi(x) = \frac{1}{2}\left\{1 + \tanh\left(\frac{x - x_c}{x_0}\right)\right\}, \qquad (4)$$

where $x_c$ and $x_0$ are fitting parameters and $x$ is the coordinate along the arrow. In this case, $x_c$, which represents the position of the SCL-CO interface, increases at a constant rate (the inset of Fig. 4a), suggesting that the interfacial reaction is the rate-controlling process. The slope of $x_c$ vs. $t$ defines the growth rate $v$, which is not appreciably dependent on the growth directions (Supplementary Note V). We repeated these experiments and analyses at different temperatures to reveal the temperature dependence of $v$, the result of which is displayed in Fig. 4b. Notably, $v$ increases roughly linearly with $\Delta T = T_{CO} - T$ near $T_{CO}$ (the inset of Fig. 4b) and is saturated to a constant value of ~$10^3$ nm/s at low temperatures. In general, the temperature variation of $v$ is discussed as follows[30,31]:

$$v = k(T)\{1 - \exp(-\Delta\mu / k_B T)\}, \qquad (5)$$

where $k(T)$ is the temperature-dependent kinetic factor for crystal growth, $\Delta\mu = \mu_{SCL} - \mu_{CO}$ is the free energy difference between the SCL and CO, and $k_B$ is the Boltzmann constant.

Notably, $\Delta\mu$ is proportional to $\Delta T$, which explains the $\Delta T$-linear variation of $v$ near $T_{CO}$.

Crystal growth in classical particle systems is generally described by the Wilson-Frenkel model[32], where $k(T)$ is proportional to the diffusion constant $D(T)$ in the supercooled liquid. Regarding electronic systems, the characteristic frequency of charge fluctuations $f(T)$ in SCL can be used instead of $D(T)$. Then, Eq. (5) becomes:

$$v = f(T)l\{1 - \exp(-\Delta\mu/k_B T)\}, \qquad (6)$$

where $l$ is the length of one crystal growth step. As in the simplest case, we take a lattice constant of θ-RbZn (0.5 nm) as $l$ and exploit the previously obtained $f(T)$ by noise spectroscopy[6] (Supplementary Note III). Since $\Delta\mu = \Delta H \Delta T / T_{CO}$ in which $\Delta H$ is the enthalpy difference between the SCL and CO (~160 K)[33], substitution of the experimental values of $l$, $f(T)$ and $\Delta H$[33] into Eq. (6) gives the blue line in Fig. 4b; however, these values are 3-5 orders of magnitude below the experimental values. Note that this difference of several orders of magnitude at low temperatures is not affected by the choice of the value of $\Delta\mu$, namely $\Delta H$.

A possible modification of the Wilson-Frenkel model is to treat $l$ as an adjustable parameter. Figure 4c is the plot of $l = v\left[f(T)\{1 - \exp\left(-\frac{\Delta\mu}{k_B T}\right)\}\right]^{-1}$ with the experimental values of $v$, $f(T)$ and $\Delta H$. It turns out that $l$ reaches an unrealistically large value of 100 μm ($10^5$ lattice constants) at 170 K. It is noted that $l$ can be longer than the lattice constant in the avalanche-mediated crystallization of the mature glass.[34]; however, it is at most on the order of 10 lattice constants. Alternatively, one may assume that $k(T)$ is not proportional to a diffusion constant, as is the case in some classical particle systems, where $k(T)$ is gapless[35–38]. This is possibly because of the crystalline-like local structure of the liquid abutting the crystal surface [30,31]. However, the fit with $k(T)=k_0$ (the orange line in Fig. 4b) yields $k_0$=840 nm/s and $\Delta H$=4600 K, which is 30 times larger than the experimental value.

Thus, the experimental growth rates are not reconciled with the conventional models. This discrepancy likely stems from the quantum nature of electrons inherited by the CG or SCL, which has moderate electrical conductivity, e.g., 10 S/cm at 150 K, that is suggestive of a spatially spreading wave function[7] and contrasts to the highly insulating CO state with the wave function well localized on each molecule. In classical systems, crystallization proceeds by the spatial rearrangement of particles; however, crystallization from the CG or SCL should be accompanied by shrinking of the wave functions and a drastic change in the quantum nature of electrons, which results in a novel case of quantum crystallization. Kinetic factor $k(T)$ values that are orders of magnitude larger than those in the conventional case are known in the quantum crystallization of liquid He[39,40]. Electronic and He crystallizations share anomalous nonequilibrium dynamics, which are attributable to their quantum nature.

In the present study, we succeeded in directly observing the spaciotemporal profile of electronic crystallization, evidencing distinct crystallization features at high and low temperatures that is consistent with the celebrated nucleation and growth mechanism. On the other hand, it has been–found that the liquid (or glass) crystal interface proceeds through space orders of magnitude faster than in classical systems, indicating that the quantum nature of electrons is deeply involved in the anomalously high speed of crystal growth. The electronic state at the SCL-CO interface and its dynamics should hold the key to the microscopic mechanism of the fast growth. The present results are expected to open a new avenue to the physics of crystallization in quantum systems and trigger theoretical studies on this issue.

## References

1. Debenedetti, P. G. & Stillinger, F. H. Supercooled liquids and the glass transition. *Nature* **410**, 259–267 (2001).

2. Debenedetti, P. G. *Metastable Liquids: Concepts and Principles*. (Princeton University Press, 1996).

3. Lubchenko, V. & Wolynes, P. G. Theory of structural glasses and supercooled liquids. *Annu. Rev. Phys. Chem.* **58**, 235–266 (2007).

4. Gasser, U. Crystallization in three-and two-dimensional colloidal suspensions. *J. Phys. Condens. Matter* **21**, (2009).

5. Strobl, G. R. *The Physics of Polymers*. (Springer, 1997).

6. Kagawa, F. *et al.* Charge-cluster glass in an organic conductor. *Nat. Phys.* **9**, 419–422 (2013).

7. Sato, T., Miyagawa, K., Tamura, M. & Kanoda, K. Anomalous 2D-Confined Electronic Transport in Layered Organic Charge-Glass Systems. *Phys. Rev. Lett.* **125**, 146601 (2020).

8. Sato, T., Miyagawa, K. & Kanoda, K. Electronic crystal growth. *Science* **357**, 1378–1381 (2017).

9. Dagotto, E. Complexity in Strongly Correlated Electronic Systems. *Science* **309**, 257–262 (2005).

10. Emery, V. J. & Kivelson, S. A. Frustrated electronic phase separation and high-temperature superconductors. *Phys. C Supercond. its Appl.* **209**, 597–621 (1993).

11. Seo, H. Charge ordering in organic ET compounds. *J. Phys. Soc. Jpn.* **69**, 805–820 (2000).

12. Schmalian, J. & Wolynes, P. G. Stripe glasses: self-generated randomness in a uniformly frustrated system. *Phys. Rev. Lett.* **85**, 836–839 (2000).

13. Mahmoudian, S., Rademaker, L., Ralko, A., Fratini, S. & Dobrosavljević, V. Glassy Dynamics in Geometrically Frustrated Coulomb Liquids without Disorder. *Phys. Rev. Lett.* **115**, 1–5 (2015).

14. Sato, T. *et al.* Emergence of nonequilibrium charge dynamics in a charge-cluster glass. *Phys. Rev. B* **89**, 1–5 (2014).

15. Sato, T. *et al.* Systematic Variations in the Charge-Glass-Forming Ability of Geometrically Frustrated θ-(BEDT-TTF)$_2$X Organic Conductors. *J. Phys. Soc. Jpn.* **83**, 083602 (2014).

16. Müller, M., Strack, P. & Sachdev, S. Quantum charge glasses of itinerant fermions with cavity-mediated long-range interactions. *Phys. Rev. A* **86**, 1–16 (2012).

17. Mori, H., Tanaka, S. & Mori, T. Systematic study of the electronic state in θ-type BEDT-TTF organic conductors by changing the electronic correlation. *Phys. Rev. B* **57**, 12023–12029 (1998).

18. Miyagawa, K., Kawamoto, A. & Kanoda, K. Charge ordering in a quasi-two-dimensional organic conductor. *Phys. Rev. B* **62**, R7679–R7682 (2000).

19. Yamamoto, K., Yakushi, K., Miyagawa, K., Kanoda, K. & Kawamoto, A. Charge ordering in θ-(BEDT−TTF)$_2$RbZn(SCN)$_4$ studied by vibrational spectroscopy. *Phys. Rev. B* **65**, 085110 (2002).

20. Watanabe, M., Noda, Y., Nogami, Y. & Mori, H. Investigation of X-ray diffuse scattering in θ-(BEDT-TTF)$_2$RbM'(SCN)$_4$. *Synth. Met.* **135–136**, 665–666 (2003).

21. Sasaki, S. *et al.* Crystallization and vitrification of electrons in a glass-forming charge liquid. *Science* **357**, 1381–1385 (2017).

22. Ouyang, J., Yakushi, K., Misaki, Y. & Tanaka, K. Raman spectroscopic evidence for the charge disproportionation in a quasi-two-dimensional organic conductor

θ-$(BDT\text{-}TTP)_2Cu(NCS)_2$. *Phys. Rev. B* **63**, 054301 (2001).

23. Drozdova, O. *et al.* Raman spectroscopy as a method of determination of the charge on BO in its complexes. *Synth. Met.* **120**, 739–740 (2001).

24. Wang, H. H., Ferraro, J. R., Williams, J. M., Geiser, U. & Schlueter, J. A. Rapid Raman spectroscopic determination of the stoichiometry of microscopic quantities of BEDT-TTF-based organic conductors and superconductors. *J. Chem. Soc. Chem. Commun.* 1893–1894 (1994).

25. Wang, H. H., Kini, A. M. & Williams, J. M. Raman characterization of the BEDT-TTF$(ClO_4)_2$ salt. *Mol. Cryst. Liq. Cryst.* **284**, 211–221 (1996).

26. Moldenhauer, J. *et al.* FT-IR absorption spectroscopy of BEDT-TTF radical salts: charge transfer and donor-anion interaction. *Synth. Met.* **60**, 31–38 (1993).

27. Yakushi, K. Infrared and Raman Studies of Charge Ordering in Organic Conductors, BEDT-TTF Salts with Quarter-Filled Bands. *Crystals* **2**, 1291–1346 (2012).

28. Avrami, M. Kinetics of phase change. I: General theory. *J. Chem. Phys.* **7**, 1103–1112 (1939).

29. Cahn, J. W. Theory of crystal growth and interface motion in crystalline materials. *Acta Metall.* **8**, 554–562 (1960).

30. Sun, G., Xu, J. & Harrowell, P. The mechanism of the ultrafast crystal growth of pure metals from their melts. *Nat. Mater.* **17**, 881–886 (2018).

31. Gao, Q. *et al.* Fast crystal growth at ultra-low temperatures. *Nat. Mater.* (2021) doi:10.1038/s41563-021-00993-6.

32. Wilson, H. W. XX. On the velocity of solidification and viscosity of super-cooled liquids. *London, Edinburgh, Dublin Philos. Mag. J. Sci.* **50**, 238–250

(1900).

33. Takeno, T. *et al.* Mysterious charge ordering on θ-(BEDT-TTF)$_2$RbZn(SCN)$_4$. *J. Phys. Conf. Ser.* **150**, (2009).

34. Sanz, E. *et al.* Avalanches mediate crystallization in a hard-sphere glass. *Proc. Natl. Acad. Sci. U. S. A.* **111**, 75–80 (2014).

35. Broughton, J. Q., Gilmer, G. H. & Jackson, K. A. Crystallization rates of a Lennard-Jones liquid. *Phys. Rev. Lett.* **49**, 1496–1500 (1982).

36. Ashkenazy, Y. & Averback, R. S. Atomic mechanisms controlling crystallization behaviour in metals at deep undercoolings. *Epl* **79**, (2007).

37. Ashkenazy, Y. & Averback, R. S. Kinetic stages in the crystallization of deeply undercooled body-centered-cubic and face-centered-cubic metals. *Acta Mater.* **58**, 524–530 (2010).

38. Orava, J. & Greer, A. L. Fast and slow crystal growth kinetics in glass-forming melts. *J. Chem. Phys.* **140**, (2014).

39. Balibar, S., Alles, H. & Parshin, Y. A. The surface of helium crystals. *Rev. Mod. Phys.* **77**, 317–370 (2005).

40. Nomura, R. & Okuda, Y. Colloquium: Quantum crystallizations of $^4$He in superfluid far from equilibrium. *Rev. Mod. Phys.* **92**, 41003 (2020).

41. Iwai, S. *et al.* Photoinduced melting of a stripe-type charge-order and metallic domain formation in a layered BEDT-TTF-based organic salt. *Phys. Rev. Lett.* **98**, (2007).

**Acknowledgments.** We thank T. Sato and K. Yamamoto for their initial trial of Raman spectroscopy on electronic crystallization, which motivated the present work, and A. Ikeda, T. Kato and K. Yoshimi for their fruitful discussion. This work was supported by Japan Society for the Promotion of Science (JSPS) under Grant Numbers 18H05225, 19H01846, 20K20894, 20KK0060, 19H02579 and 21H05234. S.A. also thanks to the support from Murata Science Foundation.

**Author contributions.** K.M. prepared samples. H.M. performed experiments and analyzed as well as interpreted data with the help of S.A., T.H., K.M. and K.K. K.K designed the project. H.M. and K.K. wrote the manuscript with the input from all authors.

**Competing financial interests.** The authors declare no competing financial interests.

**Additional information**

**Supplementary information** is available for this paper at ***.

**Correspondence and requests for materials** should be addressed to K.K.

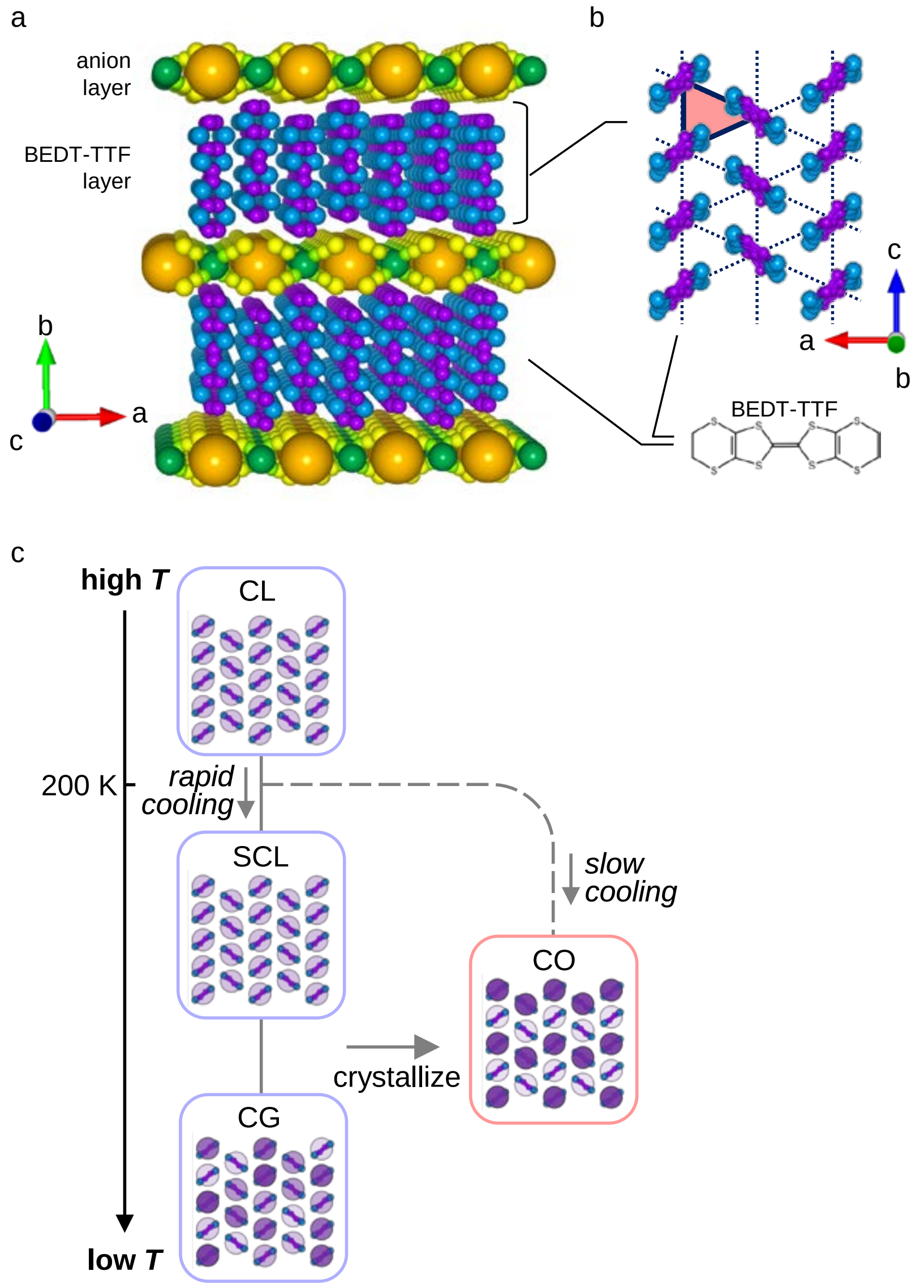

**Figure 1 | Crystal structure and electronic phases of θ-(BDET-TTF)$_2$RbZn(SCN)$_4$.**

**a**, Layered crystal structure of θ-(BEDT-TTF)$_2$RbZn(SCN)$_4$ viewed from the c-axis. The conducting BEDT-TTF layers are alternated with the insulating RbZn(SCN)$_4$ layers. The BEDT-TTF layers host charge order, charge glass, etc. **b**, In-plane structure of the BEDT-TTF layer viewed from the *b*-axis. The BEDT-TTF molecules form an anisotropic triangular lattice (highlighted by red colour) with one hole per two molecular sites. **c**, Schematic diagram of the temperature and cooling-rate dependence of the electronic phases in θ-(BEDT-TTF)$_2$RbZn(SCN)$_4$. CO, CL, SCL and CG represent the charge order, charge liquid, supercooled charge liquid and charge glass, respectively.

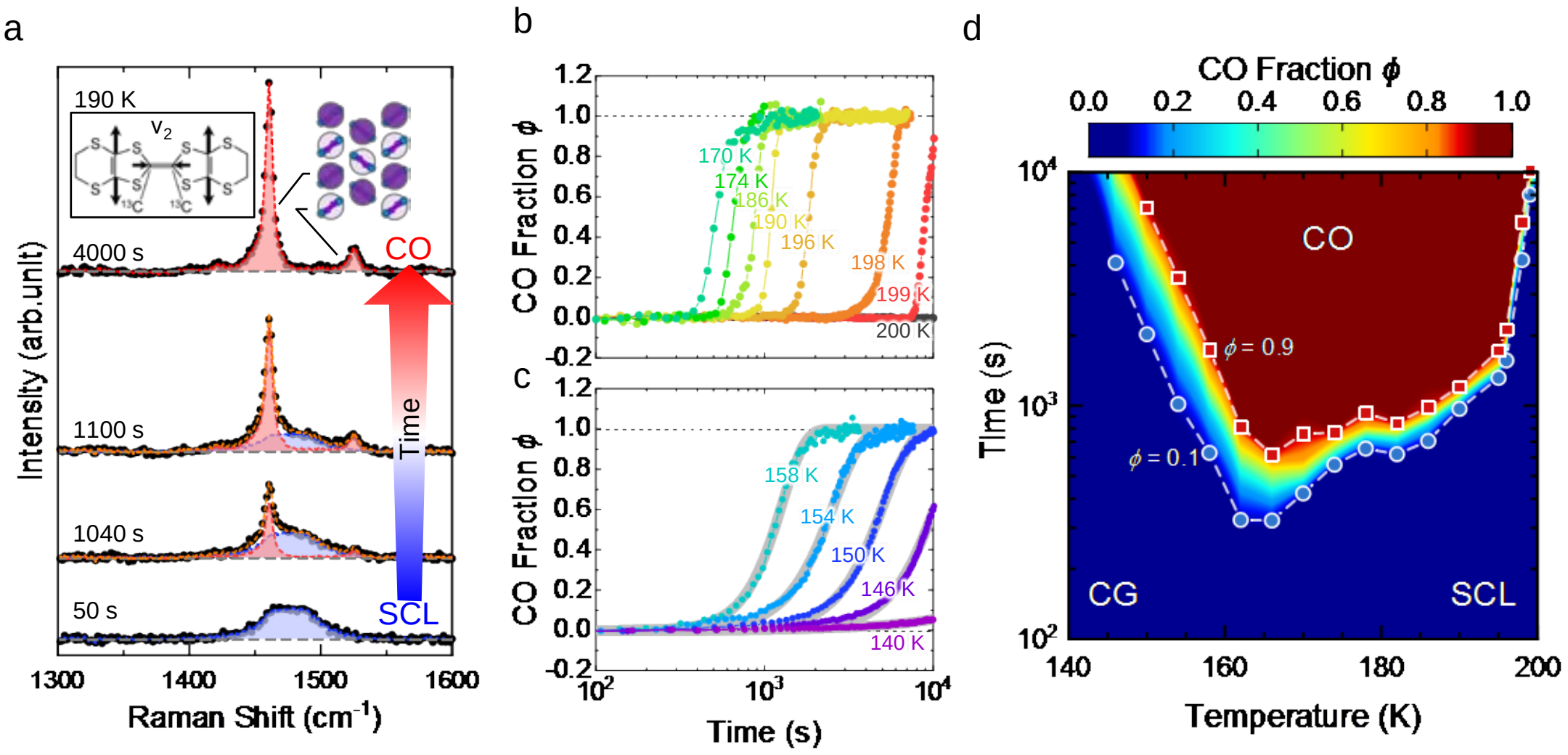

**Figure 2 | Time evolution of the Raman spectra and CO fraction during electronic crystallization and the time-temperature-transformation (TTT) diagram.**

**a**, Typical time evolution of the Raman spectra of the $\nu_2$ mode during the change from SCL/CG to CO. The excitation light polarized parallel to the a-axis is irradiated in a direction normal to the (010) surface, and the scattered light polarized parallel to the a-axis and c-axis is collected. Immediately after rapid cooling, the Raman spectrum of SCL/CG (blue) is observed. Then, at a certain time, the Raman peaks of CO (red component) emerge, and its fraction gradually increases to finally reach 100%. Inset: Charge sensitive $\nu_2$ mode of the BEDT-TTF molecule with the central double-bonded carbons enriched by $^{13}$C isotopes. **b, c**, Time evolution of the CO volume fraction $\phi$ at temperatures above (a) and below (b) the nose temperature. The grey lines are the fits of the Avrami equation (Eq. (3)) to the experimental points. **d**, Contour plot of the $\phi$ values in the time-temperature-transformation plane (TTT diagram). The circles and squares indicate the contour points of $\phi$ = 0.1 and 0.9, respectively; the dashed lines are guides for the eye. In the blue and brown areas, the sample is entirely occupied by CO and CG/SCL, respectively.

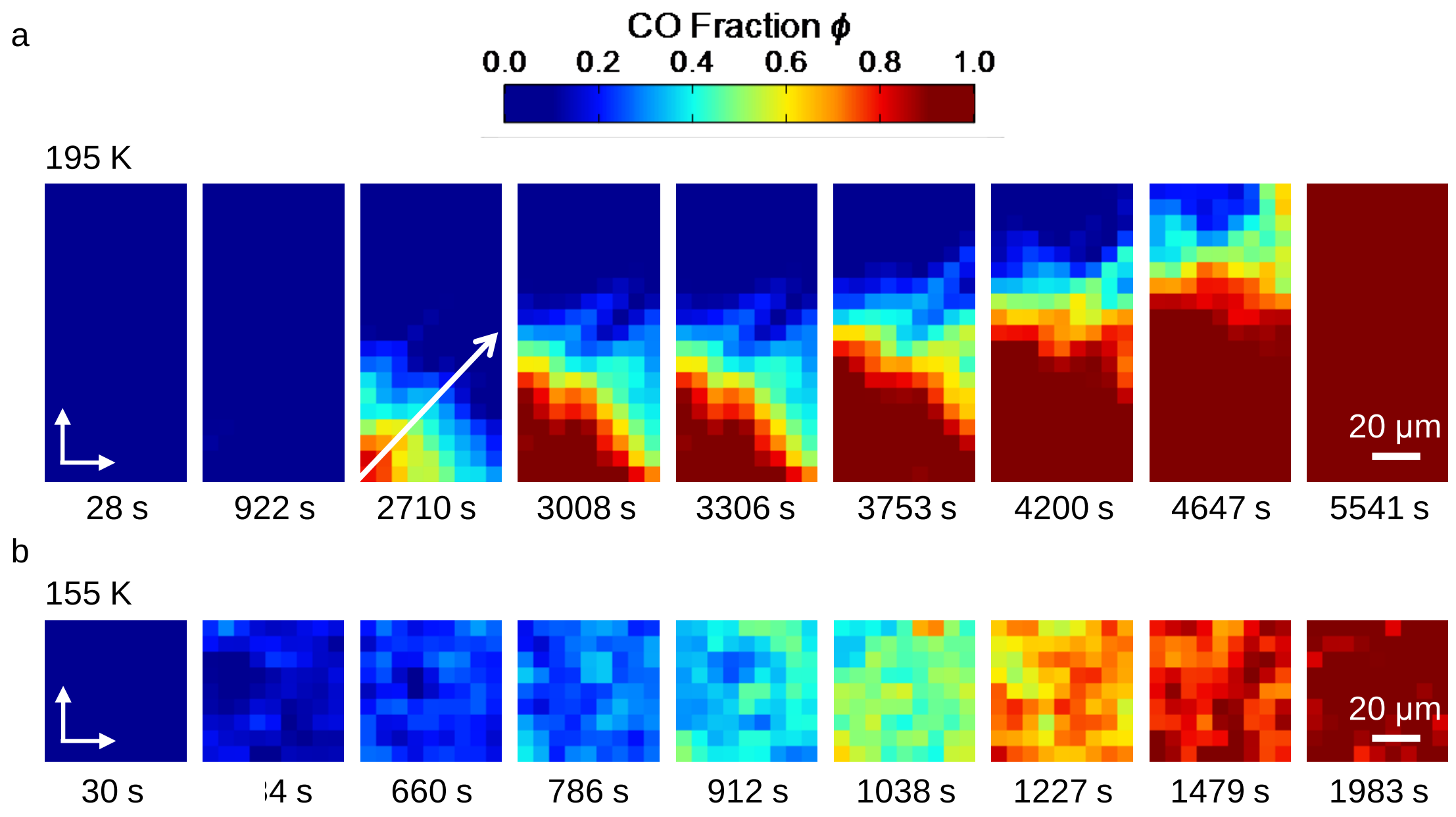

**Figure 3 | Raman images of electronic crystal growth.**

**a**, Time evolution of the real-space Raman images (contour plots of $\phi$ on the sample ac plane) at 195 K. The images shown are those taken immediately after quenching to 195 K and at 922 s, 2710 s, 3008 s, 3306 s, 3753 s, 4200 s, 4647 s and 5541 s. The white arrow in the third image is the path, along which the growth rate is determined as shown in Fig. 4a. **b**, Time evolution of the real-space Raman images at 155 K. The images shown are those taken immediately after quenching to 155 K and at 534 s, 660 s, 786 s, 912 s, 1038 s, 1227 s, 1479 s and 1983 s. We took more images with narrower time intervals at 195 K and 155 K (see the Supplementary Note II for the images at other times). The size of one pixel is $6.5 \times 6.5$ $\mu m^2$, which is the spatial resolution of the present Raman imaging instrument.

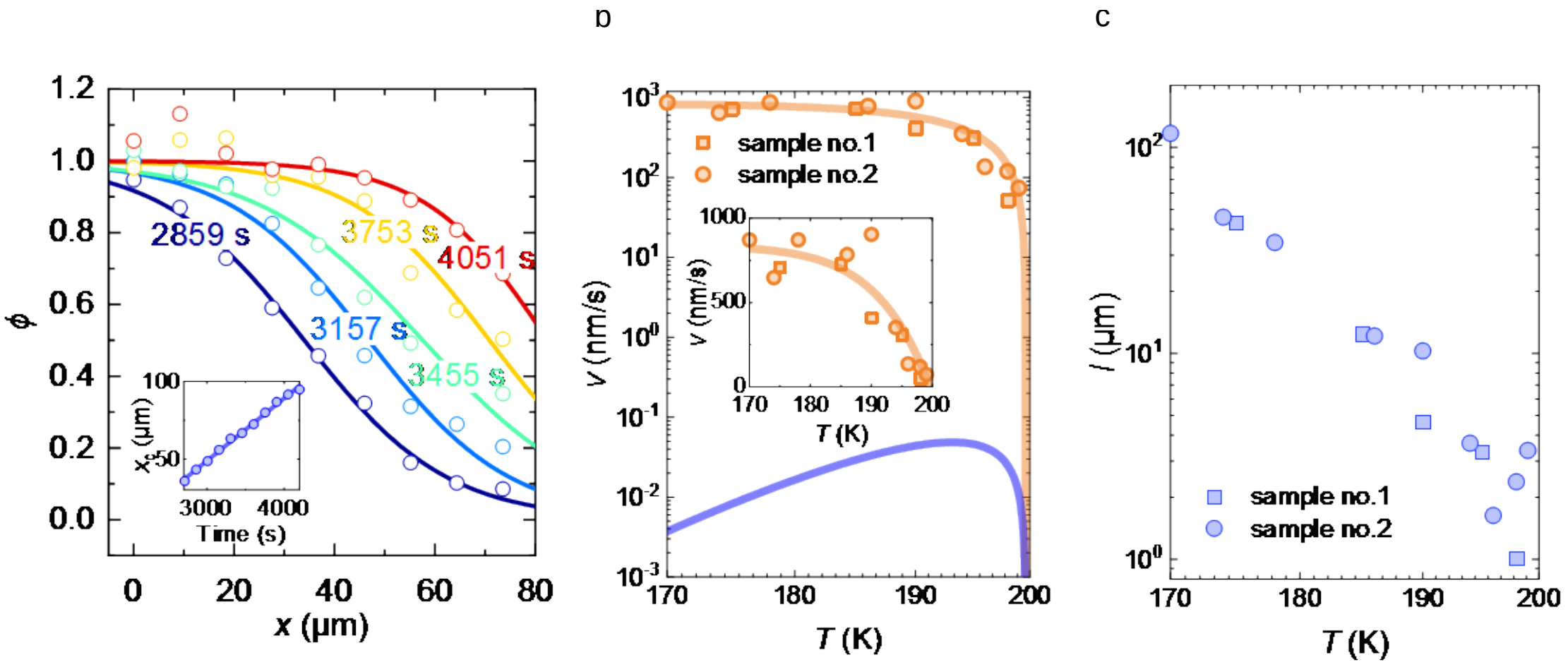

**Figure 4 | Electronic crystal growth rate.**

**a**, $\phi$ values along the arrow in Fig. 3a at various times. The solid curves are fits of Eq. (4) to the data points. Inset: Time dependence of the location of the domain boundary $x_c$. **b**, Temperature dependence of the experimental growth rate and theoretical curves. The blue line indicates the calculation based on the Wilson-Frenkel model. The orange line is a theoretical curve based on Eq. (5) with $k(T)=k_0$. **c**, Temperature dependence of the growth step length $l$ projected by the Wilson-Frenkel model with the experimental values of $v$.

## Methods

The Raman spectrum of the C=C stretching mode ($\nu_2$ mode) in BEDT-TTF, which is particularly charge-sensitive,[23-26] was exploited to distinguish between CO and SCL/CG, and their spatial distribution and time evolution were visualized by Raman microspectroscopy. We used $^{13}$C-enriched single crystals of θ-RbZn, which were synthesized by the electrochemical oxidation method, because the assignment of Raman peaks was straightforward owing to resolving of the degeneracy of the C=C stretching modes by site-selective substitution[19]. The typical crystal size was approximately $1 \times 0.1 \times 0.1$ mm$^3$. The crystals were mounted on copper substrates and then loaded on a cooling stage (Linkam 10002L), which had a glass window for optical observation. We cooled the sample at a rate of 30 K/min to metastabilize the SCL and CG; in other cases, cooling was performed at 1 K/min or more slowly. Raman spectra were measured by a Renishaw inVia Raman system. Excitation was provided from a 532 nm laser focused through a microscope equipped with a Leica N Plan L50x objective lens. The scattered light in the backscattering geometry was split by a diffraction grating of 1800 g/mm and detected by a charge coupling device. The Renishaw StreamLine technique was used for rapid Raman imaging. The investigated temperature range was 140-200 K.

The observation depth is given by the smaller of the following two lengths. One is the depth resolution of the microscope. The formula of full width at half minimum for a confocal microscope is as follows:

$$0.64 \times \frac{\lambda}{n - \sqrt{n^2 - \mathrm{NA}^2}} ,$$

where $\lambda$ is a wavelength (532 nm), $n$ is a refractive index (1) and NA is a numerical aperture of an objective lens (0.5). The substitution of these values yields the depth resolution of 2.5 μm. The other is the penetration depth of the light, given as the inverse of the absorption coefficient. According to S. Iwai *et al*[41], the absorption coefficient of θ-RbZn is 15000 $cm^{-1}$, which means that the penetration depth is 0.67 μm. Then, the observation depth turns out to be about 1 μm.

# SUPPLEMENTARY INFORATION

**Note I. Analysis of the Raman spectra**

To characterize the electronic crystallization over time in θ-RbZn, we pursued the time evolution of the Raman spectrum, which was decomposed into supercooled liquid/glass (SCL/CG) and charge order (CO) components; the time evolution of the CO fraction was examined at each temperature studied. Figure S1 shows the basis spectra of SCL/CG and CO at each temperature used for the analyses shown in Fig. 2 in the main text. $I^{CG}(\nu)$ is the Raman spectrum of SCL/CG measured immediately after rapid cooling to each temperature, and $I^{CO}(\nu)$ is the Raman spectrum of CO measured after crystallization is complete. In the measurements, a background trend that inevitably arises from stray light was superposed on the Raman spectra. As it is well approximated by the linear trend $(a\nu + b)$ in Fig. S2, we determined $a$ and $b$ by its fit to the experimental baseline outside the peak(s) (from 1300 cm$^{-1}$ to 1400 cm$^{-1}$ and from 1550 cm$^{-1}$ to 1600 cm$^{-1}$). Then, the Raman spectra $I(\nu、t)$ during crystallization are expressed as:

$$I(\nu、t) = A^{CO}(t)I^{CO}(\nu) + A^{CG}(t)I^{CG}(\nu) + (a\nu + b), \qquad (S1)$$

where $A^{CO}(t)$ and $A^{CG}(t)$ are the spectral weights of CO and SCL/CG at time $t$. The fraction of CO is evaluated by $\phi(t) = A^{CO}(t)/(A^{CO}(t) + A^{CG}(t))$. Figure S2 shows examples of the fitting at 198 K and 154 K. Figure 2a and Fig. S1 show the spectra after subtracting the background trend $(a\nu + b)$.

**Note II. Raman imaging**

Figure S3 shows all the contour plots of $\phi$ measured during the isothermal crystallization of electrons at $T_q$ = 195 K and 155 K; the contour plots in Fig. 3 in the main text are representative plots. The time intervals are 149 s for the measurements at 195 K (Fig. S3a) and 63 s for those at 155 K (Fig. S3b). The spatial resolution is 6.5 μm.

**Note III. Charge fluctuation frequency used in the analysis of the CO growth rate**

The previous study on the electrical resistance noise in θ-RbZn[1] succeeded in finding the lower and upper cut-off frequencies $f_{c1}$ and $f_{c2}$ of the charge fluctuations in the CL phase by analysing the noise spectrum with the distributed Lorentzian model. In the present analysis, as the charge fluctuation frequency, we adopted the geometric mean of $f_{c1}$ and $f_{c2}$, $f_0=(f_{c1}f_{c2})^{1/2}$. However, in the SCL/CG phase, $f_{c1}$ and $f_{c2}$ and therefore $f_0$ were unknown because accurate noise spectroscopy was hindered by crystallization. Thus, we extrapolated the $f_0$ values known in the CL phase to the SCL/CG phase at lower temperatures, assuming the activation type of the temperature dependence:

$$f_0(T) = f' \exp\left(-\frac{E_g}{T}\right), \qquad (S2)$$

where $f'$ and $E_g$ are determined to be $f' = 6.80 \times 10^{13}$ Hz and $E_g = 5902$ K by fitting Eq. S2 to the data in Fig. S4.

**Note IV. Observation of crystallization in large areas**

In Figs. 3a and 3b, the observation area is limited to $65 \times 130$ μm$^2$ and $65 \times 65$ μm$^2$, respectively, to obtain the high spatial resolution (6.5 μm) in the present apparatus. To ensure that the present observation is not specific to particular regions such as the vicinity of the domain boundaries, we also performed Raman imaging experiments for a larger area of $124.8 \times 416$ μm$^2$ by reducing the spatial resolution to 10.4 μm at 198 K (Fig. S5a) and 150 K (Fig. S5b). As seen in Figs. S5a and S5b, the results are essentially the same as shown in Figs. 3a and 3b, respectively, confirming that the observed features are not dependent of the observation area.

**Note V. Directional dependence of the growth speed**

In Fig. 4a, we presented the growth rate in a certain direction. In actuality, we examined the growth speed in different directions. Figure S6 shows the angular dependence of the growth speed obtained by varying the angle of the arrow and fixing its origin in Fig. 4a, indicating that the growth speed has almost no angular dependence.

**Note VI. About the evolution of Raman image while taking one image at high temperatures**

Here, we quantitatively estimate the time evolution of the Raman image while taking one image. From the analysis (Fig. 4a) of the time evolution of the Raman image (Fig. 3a), we know that the growth speed of the CO domain is 40 nm/s at 195 K. On the other hand, it takes 149 s to obtain one image. From these values, the growth length of the CO domain while taking one image is 6.0 μm, which is smaller than the spatial resolution, 6.5 μm. The change of the Raman image while imaging is, therefore, negligible.

**Note VII. Origin of the mottled colours in Fig. 3b.**

The images in Fig. 3b show spatially mottled colours. To characterize the pixel-by-pixel variation of the experimental $\phi$ values, we calculated the variance of $\phi$, $\sigma^2$, for the data in Fig. 3b and it is plotted as a function of the averaged $\phi$ value, $\langle\phi\rangle$, in Fig. S7a. Conceivable origins of the variance of $\phi$ are the spatial inhomogeneity in the distribution of the microcrystals and the noise in spectra.

First, we discuss the case of the inhomogeneity in the spatial distribution of microcrystals. For simplicity, we consider a site-percolation model where each site is either glass or crystal. When each site is crystal with probability, $\phi$, the expected value and the variance of the volume fraction of the crystal are $\phi$ and $\sigma_{sp}^2 = \phi(1-\phi)/N$, respectively, where $N$ is the number of sites in one pixel. Fig. S7b shows the $\langle\phi\rangle$ dependence of $\sigma_{sp}$ with $N$=10, 30, 100, which correspond to the sizes of the site, 2.1×2.1, $1.2\times1.2$ and $0.65\times0.65$ μm$^2$, respectively, in the present pixel size of $6.5\times6.5$ μm$^2$. The $\sigma_{sp}$ vanishes at $\langle\phi\rangle$=0 and 1, taking a peak at $\langle\phi\rangle$=0.5. The peak formation at $\langle\phi\rangle$=0.5 becomes suppressed as $N$ increases. This $\langle\phi\rangle$ dependence of $\sigma_{sp}$ is totally different from the experimental observation shown in Fig. 7a.

Next, we discuss the case in which the noise in spectra causes pixel-by-pixel fluctuations of $\phi$. In order to simulate this situation, we performed the following numerical calculations. First, we prepare ideal Raman spectra of CO and CG with small noise. The two spectra are added together so that the volume fraction of CO is equal to $\langle\phi\rangle$, and then the Gaussian noise is artificially added to it. In this way, 10000 spectra are calculated for each $\langle\phi\rangle$ and these spectra are fitted by the method described in the text. The variance of $\phi$,$\sigma_n^2$ ,is calculated from the variation of the fitting results. The $\langle\phi\rangle$ dependence of $\sigma_n$ is shown in Fig. S7c, where $\sigma_n$ increases monotonically with respect to $\langle\phi\rangle$. This reproduces the experimental results very well (Fig. S7d). Thus, the main origin of the mottled colours in Fig. 3b is concluded to be the noise in the Raman spectra.

1. Kagawa, F. *et al.* Charge-cluster glass in an organic conductor. *Nat. Phys.* **9**, 419–422 (2013).

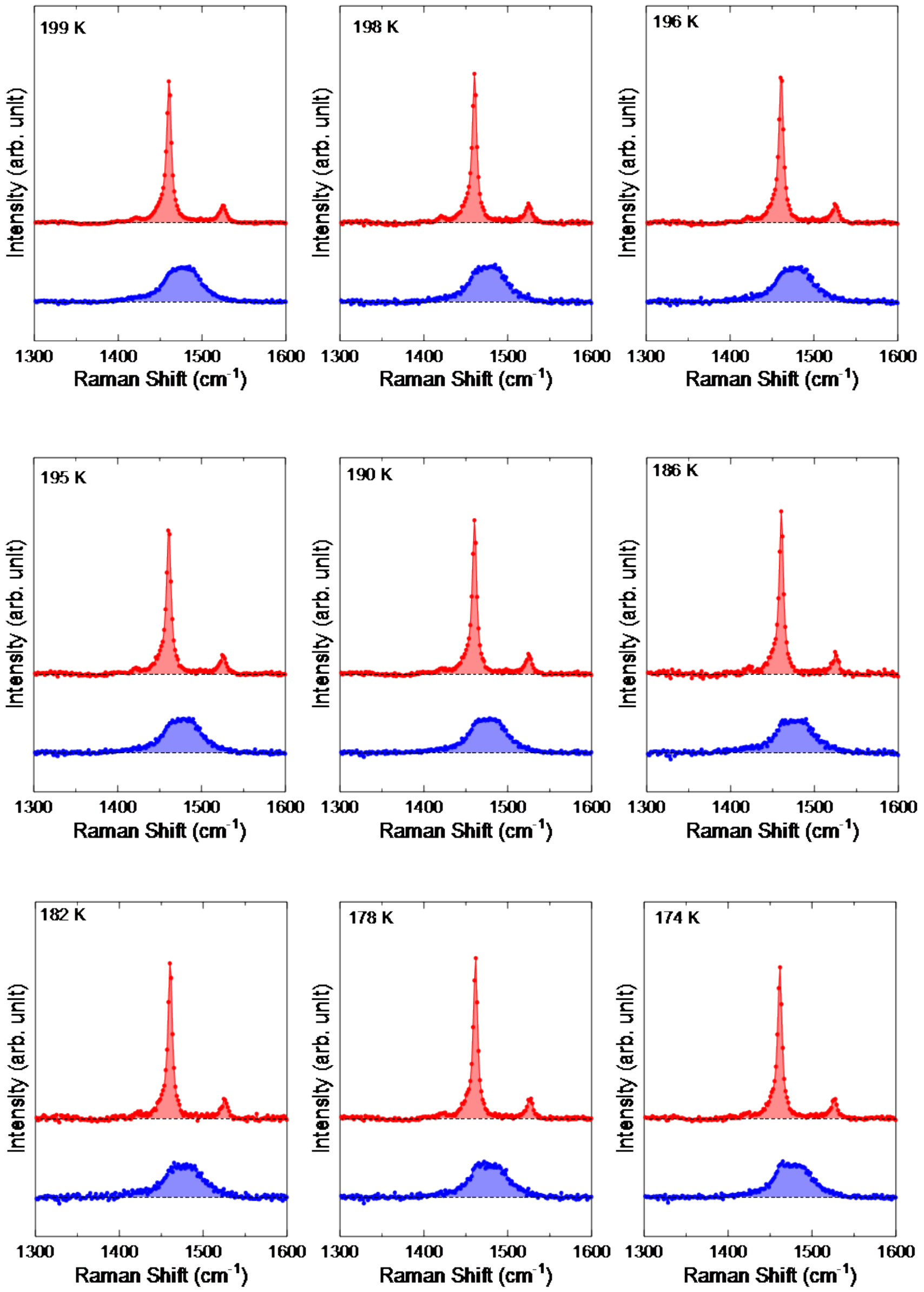

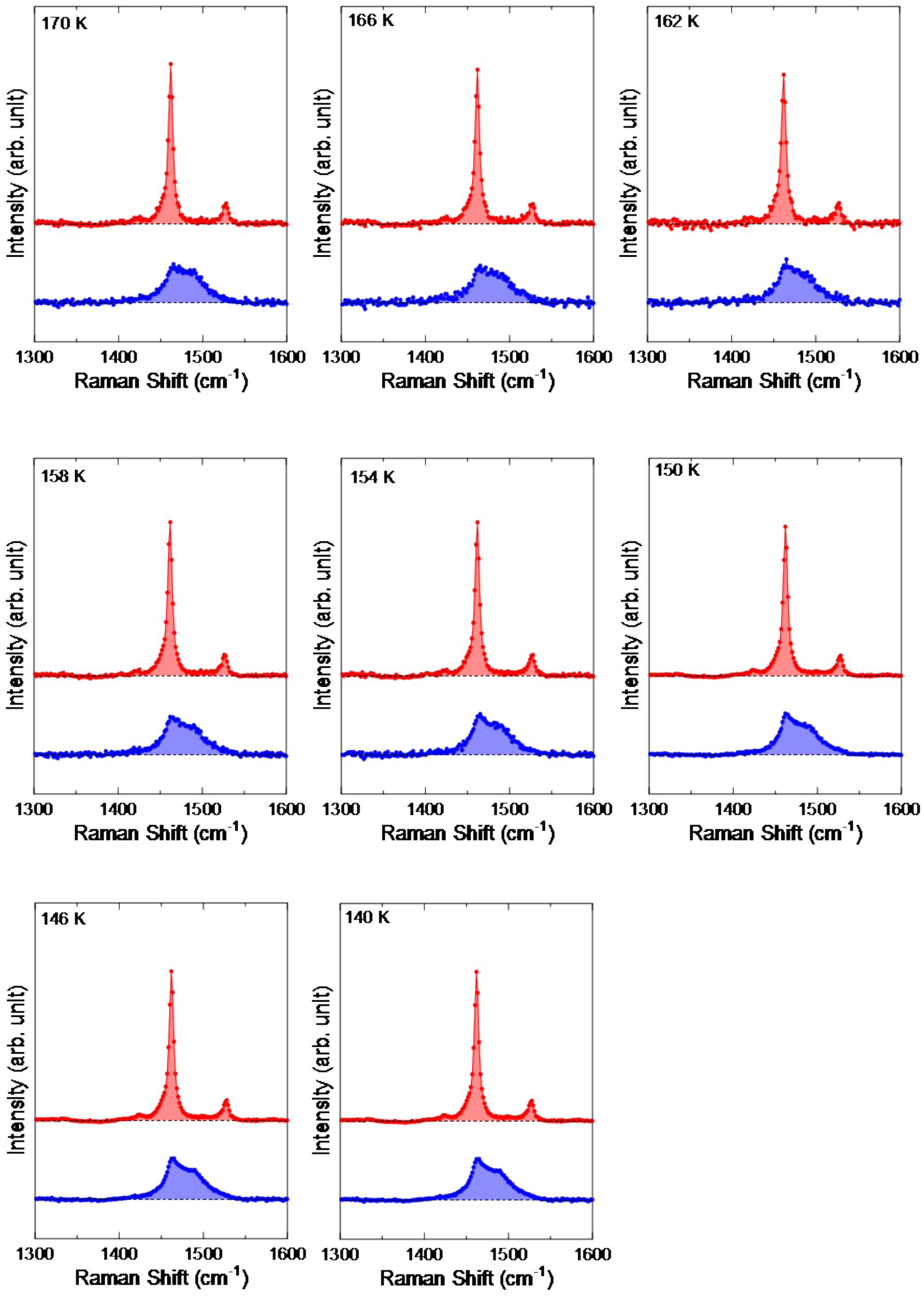

Figure S1 Raman spectra of SCL/CG (blue) and CO (red) at all temperatures studied. These spectra were used in the analysis shown in Fig. 2.

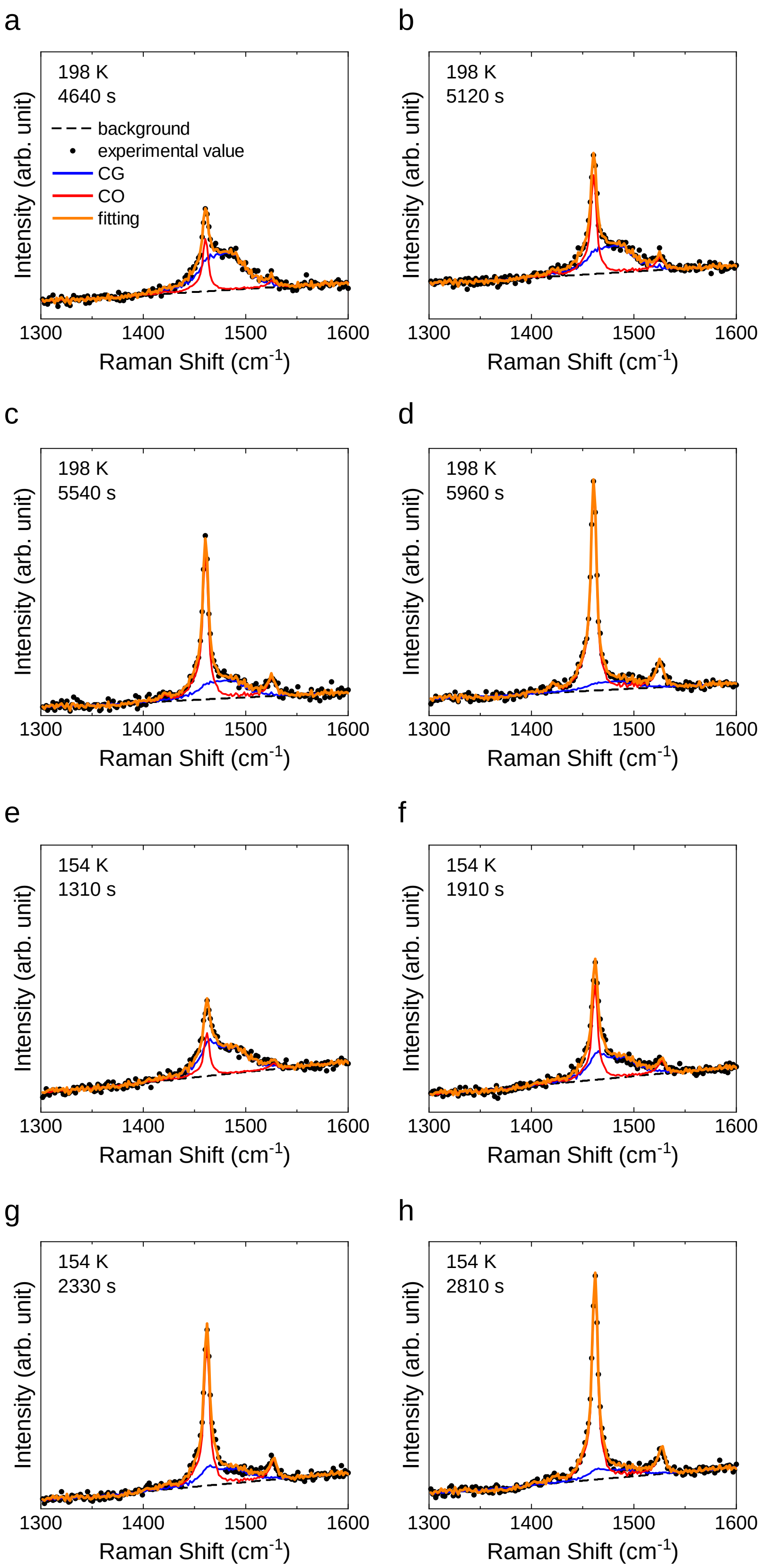

Figure S2 Raw data of the Raman spectroscopy in the crystallization process. The black, red and blue lines indicate $I(\nu, t)$ , $I^{CO}(\nu)$ and $I^{CG}(\nu)$, respectively The black dashed lines indicate background well approximated by the form of $a\nu + b$. The orange dashed lines indicate the fitting curves.

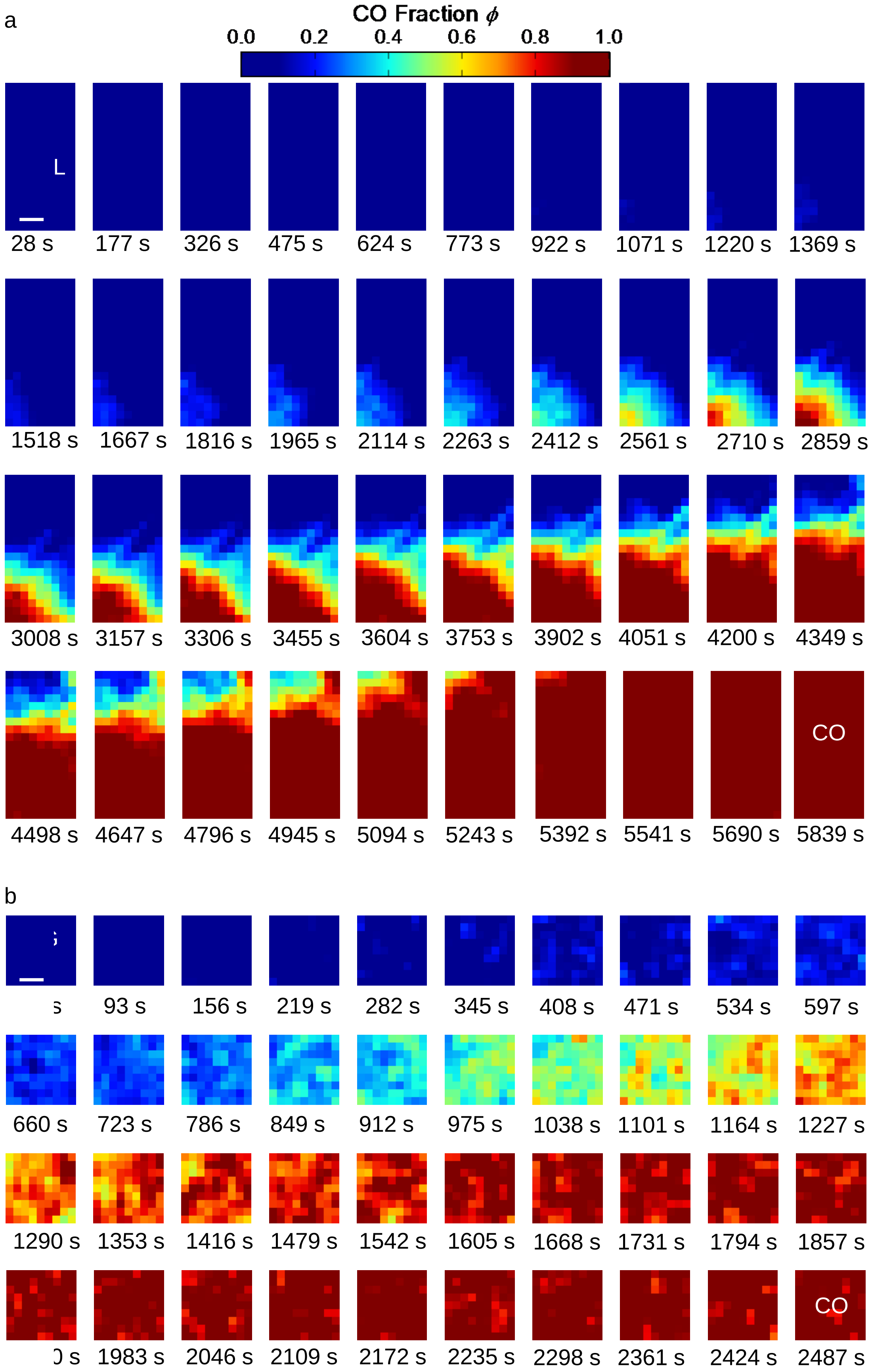

Figure S3 Time evolution of the Raman Images during isothermal crystallization. a, Crystallization at 195 K. b, Crystallization at 155 K. The scale bar indicates 20 μm.

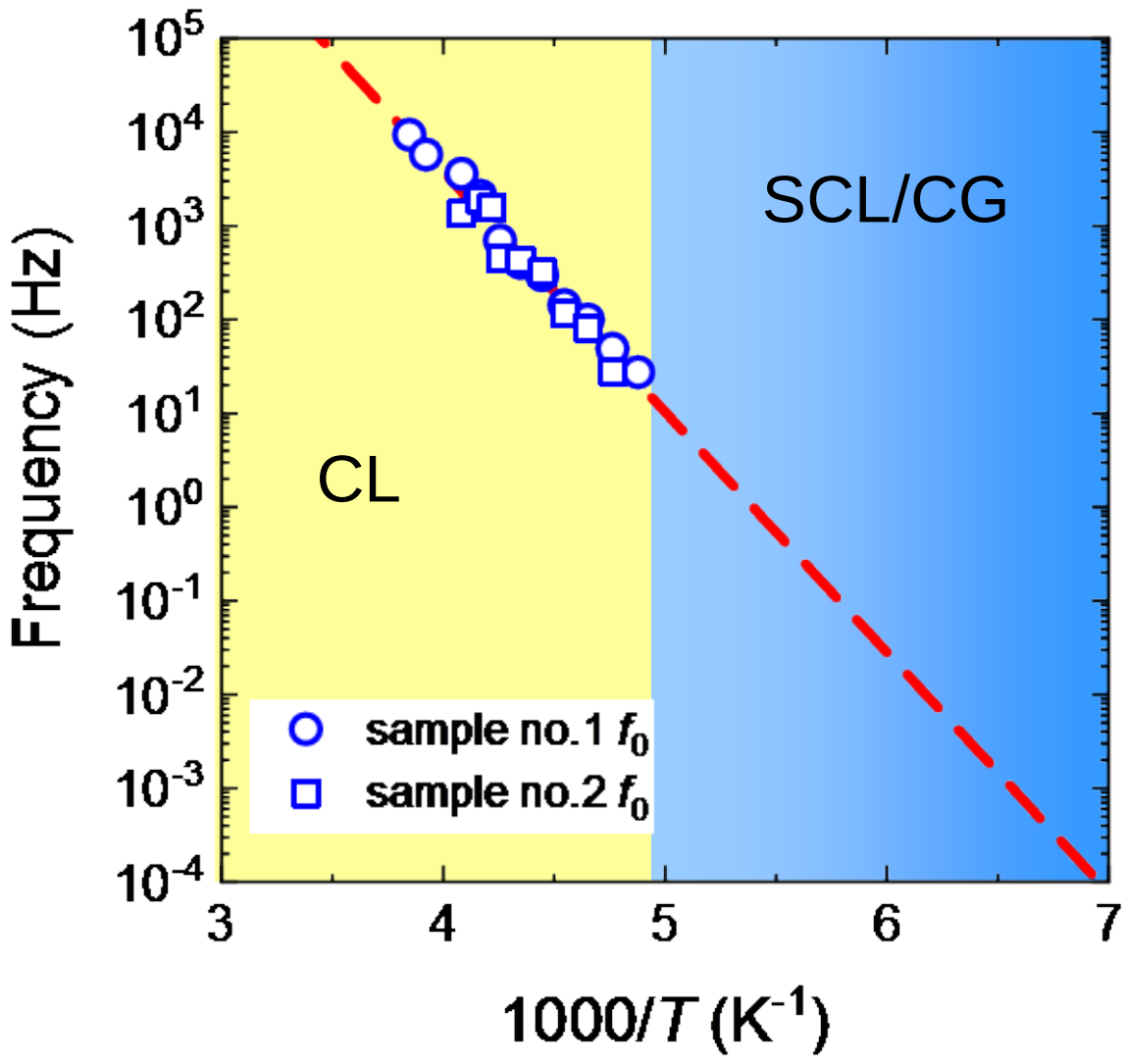

Figure S4 Extrapolation of charge fluctuation frequency. The $f_0$ [$=(f_{c1}f_{c2})^{1/2}$] values in the CL phase are plotted with using the $f_{c1}$ and $f_{c2}$ values reported in [1]. The dashed line is a fit of eq. (S2), which is extrapolated to the SCL/CG phase in the analysis (see main text).

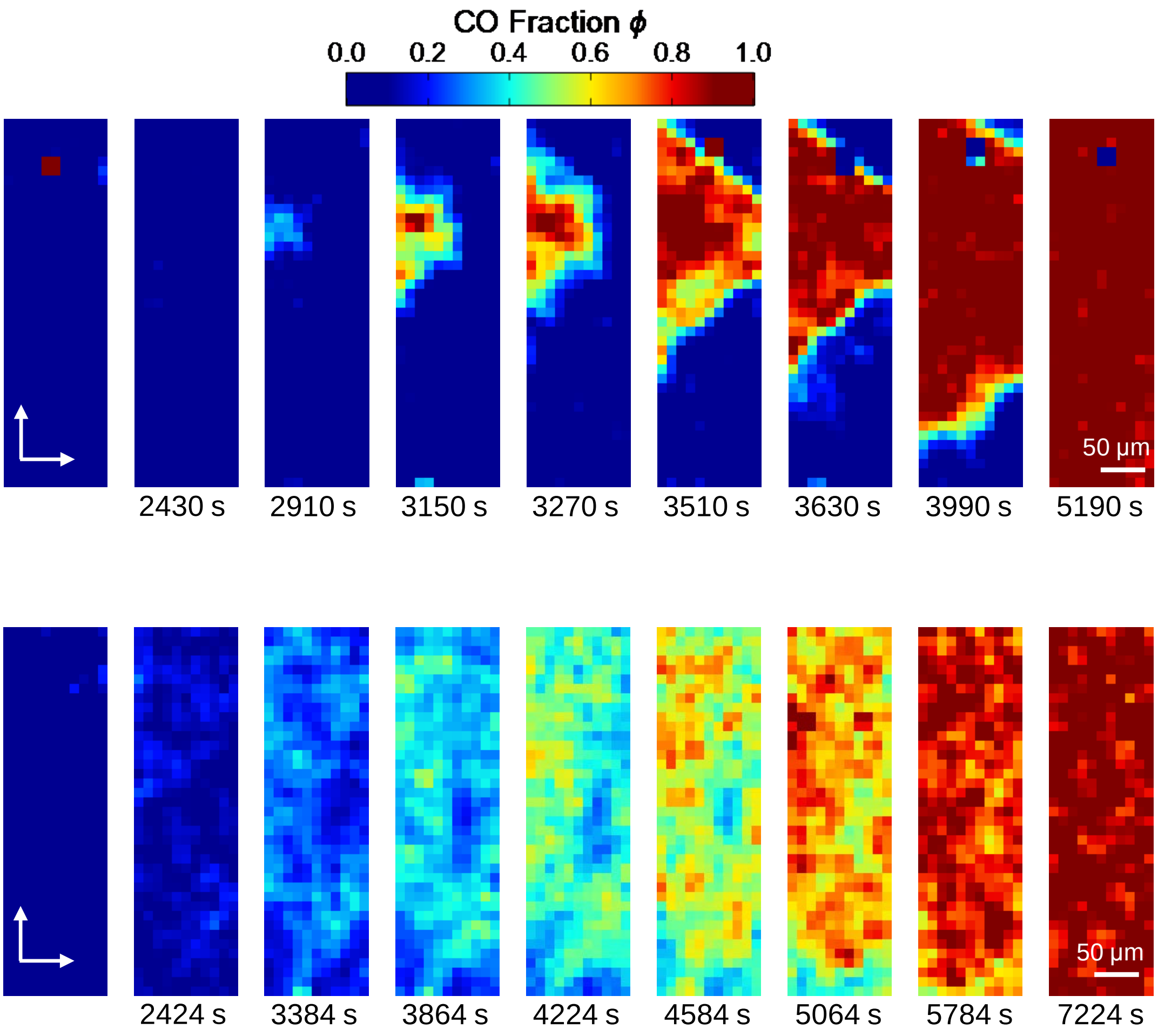

Figure S5 Time evolution of the Raman Images during isothermal crystallization in a large area. a, Crystallization at 198 K. b, Crystallization at 150 K.

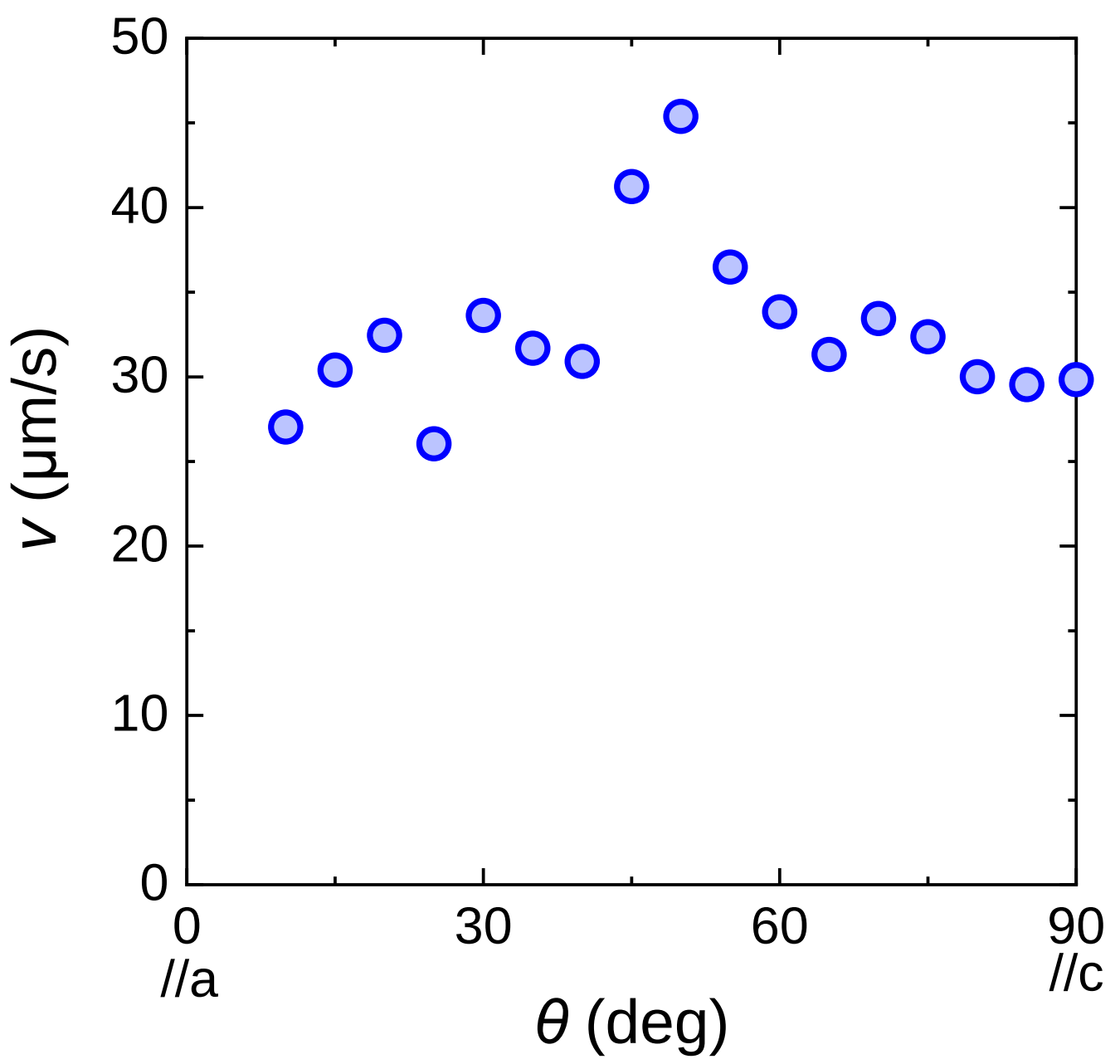

Figure S6 Directional dependence of the growth speed.

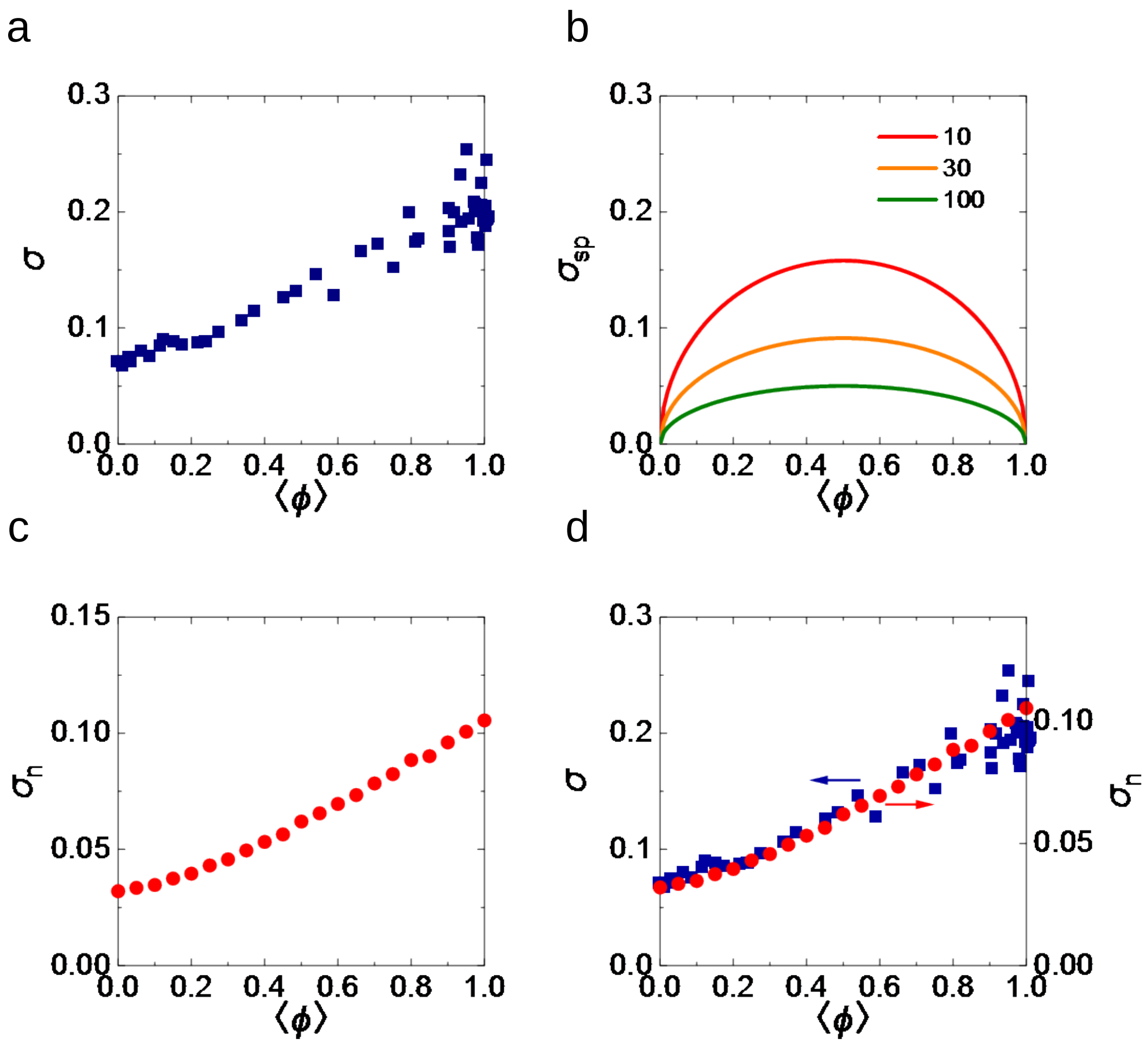

Figure S7 Experimental and simulation values of variance of $\phi$. a, experimental values of variance of $\phi$. b, Theoretical variance of $\phi$ in the site-percolation model with 10, 30 and 100 sites in one pixel. c, Calculated variance of $\phi$ for the sum of the ideal CO and CG Raman spectra with artificial gaussian noise. d, Comparison of the results of a and c.